\documentclass[12pt]{article}
\usepackage{graphicx}
\usepackage{amsmath}

\renewcommand{\theequation}{\arabic{section}.\arabic{equation}}

\begin{document}

\title{\bf Wave-function and CKM renormalization}
\author{\textsc{D. Espriu\thanks{%
espriu@ecm.ub.es}} \\
{Departament d'Estructura i Constituents de la Mat\`eria}\\
and \\
{CER for Astrophysics, Particle Physics and Cosmology} \\
{Universitat de Barcelona, Diagonal 647, Barcelona E-08028 Spain}\\
}
\date{}
\maketitle

\begin{abstract}
 In this presentation 
we clarify some aspects of the LSZ formalism and wave function
renormalization for unstable particles in the presence of electroweak
interactions when mixing and $CP$ violation are considered. We also analyze
the renormalization of the CKM mixing matrix which is closely related to
wave function renormalization. The effects due 
to the electroweak radiative corrections that are described in this work are
small, but they will need to be considered when the precision in the 
measurement of the charged
current sector couplings reaches the 1\% level.
The work presented here is done in collaboration
with Julian Manzano and Pere Talavera. 

\end{abstract}

\vfill
\vbox{
UB-ECM-PF 02/25\null\par
September 2002}

\section{Introduction}
In this conference two recent determinations of the unitarity triangle
angle $\beta$ have been presented\cite{proceedings}. BaBar finds
$\sin(2\beta)=0.75\pm 0.09\pm 0.04$, while Belle measures
$\sin(2\beta)=0.82\pm0.12\pm0.05$. The first error is statistical 
and it is rapidly decreasing, while the second one is systematical and it will
eventually limit the experimental determination
of this fundamental quantity. It is hoped that after 2007, 
LHCb\cite{Amato:1998xt}
will be able to reduce the overall uncertainty to less than 1\% .

Given this state of affairs, it is clear that in a not too distant future
a day will come 
when our experimental colleagues, when reporting on their high precision
measurement of the Kobayashi-Maskawa matrix elements, will have 
to tell us in which renormalization scheme the corresponding mixing angles 
and $CP$ violating phase(s) have 
been measured, exactly as they do so when reporting on the measured value
of $\alpha_s$. 

It is essential to have the renormalization
mechanism 
well under control to be able to separate the radiative 
corrections due to electroweak 
physics from those due to new physics. Most likely the latter  ---if 
they exist at all--- are of the
same size or even smaller than the former, exactly as a similar analysis
in the neutral current sector has shown in recent years. Effective 
lagrangian techniques are important in this context\cite{espriumanz}.

In the neutral sector it is already totally mandatory to include electroweak
radiative corrections to bring theory and experiment into agreement. Tree
level results are incompatible with experiment by many standard deviations
\cite{Pich:1997ga}. In a few years electroweak radiative corrections will be
required in the studies analysing the ``unitarity'' of the CKM matrix too
\footnote{%
The CKM matrix is certainly unitary, but the physical observables that at
tree level coincide with these matrix elements certainly do not necessarily
fulfil a unitarity constraint once quantum corrections are switched on.See
e.g. \cite{espriumanz} for a discussion on this point}.
Corrections are of several types. With an on-shell\cite{Hollik}
 scheme in mind, we
need counter terms for the electric charge, Weinberg angle and wave-function
renormalization (wfr.) for the $W$ gauge boson. We shall also require wfr.
for the external fermions and counter terms for the entries of the CKM
matrix. These are in fact related in a way that will be described below
\cite{Balzereit:1999id}. Finally one needs to compute the 1PI vertex parts
of the different processes.

Several proposals have been put forward in the literature to define
appropriate counter terms both for the external legs and for the CKM matrix
elements. The original prescription for wfr. diagonalizing the on-shell
propagator was introduced in \cite{Aoki}. In \cite{DennerSack} the wfr.
``satisfying'' the conditions of \cite{Aoki} were derived. However since
\cite{DennerSack} does not take care about the branch cuts present in the
self-energies those results must be considered only consistent up to
absorptive terms. Later it was realized \cite{Denner} that the on-shell
conditions defined in \cite{Aoki} where inconsistent and in fact impossible
to satisfy for a minimal set of renormalization constants due
to the imaginary branch cuts present in the self-energies. The author of
\cite{Denner} circumvented this problem by introducing a prescription that
\textit{de facto} eliminates such branch cuts, but at the price of not
diagonalizing the propagators in flavour space.

As we shall see later, some
Ward identities based on the SU(2)$_{L}$ gauge symmetry relate wfr. and
counter terms for the CKM matrix elements \cite{Balzereit:1999id}. 
So two of the necessary ingredients to renormalize the charged current
vertices are actually related.
In \cite
{Grassi} it was seen that if the prescription of \cite{DennerSack} was used
in the counter terms for the CKM matrix elements, the results were in
violation of gauge invariance. An
additional condition for the 
gauge invariance of the physical amplitudes is that
counter terms for the CKM matrix elements $K_{ij}$ are by themselves gauge
independent. This condition is fulfilled by the CKM counter term proposed in
\cite{Grassi} as it is in minimal subtraction \cite{Balzereit:1999id}, \cite
{Diener:2001qt}.

Other proposals to handle charged vertex  renormalization 
exist in the literature \cite{Diener:2001qt}. In all these works either
the external wfr. proposed originally in \cite{DennerSack,Denner}
are used, or the issue is sidestepped altogether. In either case the
absorptive part of the self-energies are not taken
into account. As we shall see doing so leads to $S$
-matrix elements which are gauge dependent, and this irrespective of the
method one uses to renormalize $K_{ij}$ provided the redefinition of $K_{ij}$
is gauge independent and preserves unitarity.

A more detailed account of this work is presented in \cite{EMT}

\section{W$^{+}$ and top decay}
\renewcommand{\theequation}{\arabic{section}.\arabic{equation}} %
\setcounter{equation}{0}
\label{wandtdecay}
We shall consider  
$W^{+}\left( q\right) \rightarrow f_{i}\left( p_{1}\right) \bar{f}%
_{j}\left( p_{2}\right) $ and 
$f_{i}\left( p_{1}\right) \rightarrow W^{+}\left( q\right) f_{j}\left(
p_{2}\right)$.  Latin indices
correspond to families. For the
first process there are at the one-loop level two different type
of Lorentz structures that appear
\begin{eqnarray}
M_{L}^{\left( 1\right) } &=&\bar{u}_{i}\left( p_{1}\right) \not{\varepsilon}%
\left( q\right) Lv_{j}\left( p_{2}\right) \,,\qquad \left( L\leftrightarrow
R\right) \,,  \notag \\
M_{L}^{\left( 2\right) } &=&\bar{u}_{i}\left( p_{1}\right) Lv_{j}\left(
p_{2}\right) p_{1}\cdot \varepsilon \left( q\right) \,,\qquad \left(
L\leftrightarrow R\right) \,.  \label{wdtree}
\end{eqnarray}
and for the second one
\begin{eqnarray}
M_{L}^{\left( 1\right) } &=&\bar{u}_{j}\left( p_{2}\right) \not{\varepsilon}%
^{\ast }\left( q\right) Lu_{i}\left( p_{1}\right) \,,\qquad \left(
L\leftrightarrow R\right) \,,  \notag \\
M_{L}^{\left( 2\right) } &=&\bar{u}_{j}\left( p_{2}\right) Lu_{i}\left(
p_{1}\right) p_{1}\cdot \varepsilon ^{\ast }\left( q\right) \,,\qquad \left(
L\leftrightarrow R\right) \,.  \label{tdtree}
\end{eqnarray}
At tree level 
\begin{equation}
\mathcal{M}_{0}=-\frac{eK_{ij}}{2s_{W}}M_{L}^{\left( 1\right) }\,, 
\end{equation}
where Eq. (\ref{wdtree}) is used for $M_{L}^{\left( 1\right) }$ in $W^{+}$
decay and Eq. (\ref{tdtree}) instead for $M_{L}^{\left( 1\right) }$ in $t$
decay. The one-loop corrected transition amplitude can be written as
\begin{equation}
\mathcal{M}_{1} = -\frac{e}{2s_{W}}M_{L}^{\left( 1\right) }\left[
K_{ij}\left( 1+\frac{\delta e}{e}-\frac{\delta s_{W}}{s_{W}}+\frac{1}{2}%
\delta Z_{W}\right)  
+\frac{1}{2}\left( \delta \bar{Z}%
_{ir}^{Lu}K_{rj}+K_{ir}\delta Z_{rj}^{Ld}\right) \right. \notag
\end{equation}
\begin{equation}
\left.  
 +\delta K_{ij}\right]
-\frac{e}{2s_{W}}\left( \delta F_{L}^{\left( 1\right) }M_{L}^{\left(
1\right) }+M_{L}^{\left( 2\right) }\delta F_{L}^{\left( 2\right)
}+M_{R}^{\left( 1\right) }\delta F_{R}^{\left( 1\right) }+M_{R}^{\left(
2\right) }\delta F_{R}^{\left( 2\right) }\right) .  \label{vertex}
\end{equation}
In this expression $\delta F_{L,R}^{\left( 1,2\right) }$ are the electroweak
form factors from one-loop vertex diagrams. The renormalization
constants for $e$, $s_W$ and the wfr. of the gauge boson can be found 
in \cite{Hollik}. $\delta{K_{ij}}$ and the fermion wfr. will be discussed
next.

\section{The Role of Ward Identities}

\renewcommand{\theequation}{\arabic{section}.\arabic{equation}} %
\setcounter{equation}{0}
\label{ward}

There is 
a SU(2) Ward identity 
\cite{Grassi} that relates the CKM counterterms
and wfr. constants. Let us see how the argument goes.
In the weak basis, doublets renormalize with a common wfr. constant
\begin{equation}
\left(\begin{matrix}u_0\cr d_0\cr \end{matrix}\right)_L= Z^{L \frac{1}{2}}
\left(\begin{matrix}u\cr d\cr\end{matrix}\right)_L.
\end{equation}
On the other hand, in the mass diagonal basis 
there is no reason for up-type and down-type quarks to renormalize in the
same way
\begin{equation}
\left(\begin{matrix}\tilde{u}_0\cr \tilde{d}_0\cr\end{matrix}\right)_L
= \left(\begin{matrix}Z^{uL\frac{1}{2}} u\cr Z^{dL\frac{1}{2}} d\cr 
\end{matrix}\right)_L.
\end{equation}
The passage from one basis to the other is accomplished with the help of
unitary matrices $V^0$ and $V$ for the up-type and down-type quarks, namely
\begin{equation}
\tilde{u}= V_u^\dagger u \qquad \tilde{u_0}= V_u^{0 \dagger} u_0,\qquad
\tilde{d}= V_d^\dagger u \qquad \tilde{d_0}= V_d^{0 \dagger} d_0.
\end{equation}
Elementary manipulations allow us to arrive at the 
following identity involving wfr. constants in the 
mass diagonal basis and the CKM matrix
\begin{equation} 
Z^{u L\frac{1}{2}} K = K^0 Z^{d L\frac{1}{2}},
\end{equation}
and, writing $K^0 = K + \delta K$, we arrive at
\begin{equation}
\delta K_{jk}=\frac{1}{4}\left[ \left( \delta \hat{Z}^{uL}-\delta \hat{Z}%
^{uL\dagger }\right) K-K\left( \delta \hat{Z}^{dL}-\delta \hat{Z}^{dL\dagger
}\right) \right] _{jk}\,,  \label{deltaK}
\end{equation}
where we have changed
notation and used $\hat{Z}$ 
for the wfr. constants appearing in the above expression.
Indeed, they 
are not necessarily the same ones that have to be used to renormalize and
guarantee the proper on-shell residue for the external legs and we anticipate 
that they will not. The reason is clear: the above wfr. constants
$\hat{Z}$ are introduced so as to preserve the diagonal character of 
the mass matrix and they do not necessarily keep the kinetic terms 
diagonal on-shell.

From the above expressions it is also straightforward to
derive the following  Ward identity
\begin{equation}
(\hat{Z}^{uL}\hat{Z}^{uL \dagger})^{\frac{1}{2}} K
=K (\hat{Z}^{dL}\hat{Z}^{dL \dagger})^{\frac{1}{2}}
\label{wardident}\end{equation}
Notice that in order to arrive to (\ref{deltaK}) we have used the fact
that both $K$ and $K^0$ are unitary matrices. It is perfectly possible,
though perhaps a bit strange,
to use a renormalized CKM matrix that is not unitary. If this is 
the case, the appropriate expression for the counterterm
would simply be
\begin{equation}
\delta K_{jk}=\frac{1}{2}\left[ \delta \hat{Z}^{uL}K-K
\delta \hat{Z}^{dL} \right] _{jk}\,.  \label{deltaKK}
\end{equation}
The previous Ward identity is certainly a necessary condition for the
gauge invariance of the results, but it is not 
sufficient.

Any renormalization scheme that is manifestly gauge invariant and in addition
mass independent, will obviously fulfill the above Ward identity automatically.
This can be seen explicitly from the calculations presented 
in \cite{Balzereit:1999id,Grassi} that use minimal subtraction
and a mass independent scheme, respectively. However, the 
on-shell conditions to be imposed on the external legs are manifestly 
different for different quarks, since they have different masses. It is 
therefore impossible that the external leg wfr.
 obey the previous Ward identity and they
cannot be used to define the CKM counterterms.

\section{Renormalization of External Legs}

\renewcommand{\theequation}{\arabic{section}.\arabic{equation}} 
\setcounter{equation}{0}
\label{legs}
We want to define an on-shell renormalization scheme that guarantees the
correct properties of the fermionic propagator in the $p^{2}\rightarrow
m_{i}^{2}$. The
conditions necessary for that purpose were first given by Aoki et. al. in
\cite{Aoki}. We
renormalize the bare fermion fields as
$\Psi _{0}=Z^{\frac{1}{2}}\Psi$ and  $\bar{\Psi}_{0}=\bar{\Psi}\bar{Z}^{%
\frac{1}{2}}$
For reasons that will become clear along the discussion, we shall allow $Z$
and $\bar{Z}$ to be independent renormalization constants.
Due to radiative corrections the propagator mixes fermion of different
family indices. Namely
\begin{equation}
iS^{-1}\left( p\right) =\bar{Z}^{\frac{1}{2}}\left( \frac{{}}{{}}\not{p}%
-m-\delta m-\Sigma \left( p\right) \right) Z^{\frac{1}{2}}\,,
\end{equation}
where the bare self-energy $\Sigma $ is non-diagonal and is given by $%
-i\Sigma =\sum $1PI. Within one-loop accuracy we can write $Z^{\frac{1}{2}%
}=1+\frac{1}{2}\delta Z$ etc. Introducing the family indices explicitly we
have
\begin{equation}
iS_{ij}^{-1}\left( p\right) =\left( \not{p}-m_{i}\right) \delta _{ij}-\hat{%
\Sigma}_{ij}\left( p\right) \,,
\end{equation}
where the one-loop renormalized self-energy is given by
\begin{equation}
\hat{\Sigma}_{ij}\left( p\right) =\Sigma _{ij}\left( p\right) -\frac{1}{2}%
\delta \bar{Z}_{ij}\left( \not{p}-m_{j}\right) -\frac{1}{2}\left( \not{p}%
-m_{i}\right) \delta Z_{ij}+\delta m_{i}\delta _{ij}\,.  \label{renself}
\end{equation}h
The conditions \cite{Aoki} necessary to avoid mixing will be
\begin{eqnarray}
\hat{\Sigma}_{ij}\left( p\right) u_{j}^{\left( s\right) }\left( p\right)
&=&0\,,\qquad (p^{2}\rightarrow m_{j}^{2})\,,\quad \mathrm{(incoming}\text{ }%
\mathrm{particle)}  \label{inparticle} \\
\bar{v}_{i}^{\left( s\right) }\left( -p\right) \hat{\Sigma}_{ij}\left(
p\right) &=&0\,,\qquad (p^{2}\rightarrow m_{i}^{2})\,,\quad \mathrm{(incoming%
}\text{ }\mathrm{anti}\mathrm{-}\mathrm{particle)}  \label{inantiparticle} \\
\bar{u}_{i}^{\left( s\right) }\left( p\right) \hat{\Sigma}_{ij}\left(
p\right) &=&0\,,\qquad (p^{2}\rightarrow m_{i}^{2})\,,\quad \mathrm{(outgoing%
}\text{ }\mathrm{particle)}  \label{outparticle} \\
\hat{\Sigma}_{ij}\left( p\right) v_{j}^{\left( s\right) }\left( -p\right)
&=&0\,,\qquad (p^{2}\rightarrow m_{j}^{2})\,,\quad \mathrm{(outgoing}\text{ }%
\mathrm{anti}\mathrm{-}\mathrm{particle)}  \label{outantiparticle}
\end{eqnarray}
where no summation over repeated indices is assumed and $i\neq j.$ These
relations determine the non-diagonal parts of $Z$ and $\bar{Z}$.

Here it is worth to make one important comment regarding the above
conditions. First of all we note that in the literature the relation
\begin{equation}
\bar{Z}^{\frac{1}{2}}=\gamma ^{0}Z^{\frac{1}{2}\dagger }\gamma ^{0}\,,
\label{hermiticity}
\end{equation}
is taken for granted. This relation is tacitly assumed in \cite{Aoki} and
explicitly required in \cite{Denner}. Imposing Eq. (\ref{hermiticity}) would
guarantee hermiticity of the Lagrangian written in terms of the renormalized
physical fields. However, we are at this point concerned with external leg
renormalization, for which it is perfectly possible to use a different set
of renormalization constants. In fact, using
two sets of renormalization constants is a standard practice in the on-shell
scheme \cite{Hollik}. In case one is worried about the 
consistency of using a set of wfr.
constants not satisfying (\ref{hermiticity}) for the external legs while
keeping a Hermitian Lagrangian, it should be pointed out that there is a
complete equivalence between the set of renormalization constants we shall
find out below and a treatment of the external legs where diagrams with
self-energies (including mass counter terms) are inserted instead of the
wfr. constants; provided that the mass counter term satisfy the
on-shell condition. This  gives results identical to ours
and different from those obtained using the wfr. proposed in \cite{Denner},
which do fulfil (\ref{hermiticity}). 

In any case, self-energies develop absorptive terms and this makes Eq. (\ref
{hermiticity}) incompatible with the diagonalizing conditions (\ref
{inparticle})-(\ref{outantiparticle}). Therefore in order to circumvent this
problem one can give up diagonalization conditions (\ref{inparticle})-(\ref
{outantiparticle}) or alternatively the hermiticity condition (\ref
{hermiticity}). The approach taken originally in \cite{Denner} and works
thereafter was the former alternative, while in this work we shall advocate
the second one. 

The approach of \cite{Denner} consists in dropping out
absorptive terms from conditions (\ref{inparticle})-(\ref{outantiparticle}).
Two severe problems arise if one follows this approach:
(a) Since only the dispersive part of the self-energies enters into
the diagonalizing conditions the on-shell propagator remains non-diagonal.
(b) The on-shell scheme based
in this prescription leads to gauge parameter dependent
physical amplitudes. The reason for this unwanted dependence is the dropping
of absorptive gauge parameter dependent terms in the self-energies that are
necessary to cancel absorptive terms appearing in the vertices.

We shall now present the
renormalization constants derived solely from the on-shell
conditions (\ref{inparticle})-(\ref{outantiparticle}) and allowing for $\bar{%
Z}^{\frac{1}{2}}\neq \gamma ^{0}Z^{\frac{1}{2}\dagger }\gamma ^{0}$. 
In a rather obvious notation
\begin{equation}
\delta Z_{ij}^{L} =\frac{2}{m_{j}^{2}-m_{i}^{2}}\left[ \Sigma
_{ij}^{\gamma R}\left( m_{j}^{2}\right) m_{i}m_{j}+\Sigma _{ij}^{\gamma
L}\left( m_{j}^{2}\right) m_{j}^{2}+m_{i}\Sigma _{ij}^{L}\left(
m_{j}^{2}\right) +\Sigma _{ij}^{R}\left( m_{j}^{2}\right) m_{j}\right] 
\notag
\end{equation}
\begin{equation}
\delta Z_{ij}^{R} =\frac{2}{m_{j}^{2}-m_{i}^{2}}\left[ \Sigma
_{ij}^{\gamma L}\left( m_{j}^{2}\right) m_{i}m_{j}+\Sigma _{ij}^{\gamma
R}\left( m_{j}^{2}\right) m_{j}^{2}+m_{i}\Sigma _{ij}^{R}\left(
m_{j}^{2}\right) +\Sigma _{ij}^{L}\left( m_{j}^{2}\right) m_{j}\right] 
\notag
\end{equation}
\begin{equation}
\delta \bar{Z}_{ij}^{L} =\frac{2}{m_{i}^{2}-m_{j}^{2}}\left[ \Sigma
_{ij}^{\gamma R}\left( m_{i}^{2}\right) m_{i}m_{j}+\Sigma _{ij}^{\gamma
L}\left( m_{i}^{2}\right) m_{i}^{2}+m_{i}\Sigma _{ij}^{L}\left(
m_{i}^{2}\right) +\Sigma _{ij}^{R}\left( m_{i}^{2}\right) m_{j}\right] 
\notag 
\end{equation}
\begin{equation}
\delta \bar{Z}_{ij}^{R} =\frac{2}{m_{i}^{2}-m_{j}^{2}}\left[ \Sigma
_{ij}^{\gamma L}\left( m_{i}^{2}\right) m_{i}m_{j}+\Sigma _{ij}^{\gamma
R}\left( m_{i}^{2}\right) m_{i}^{2}+m_{i}\Sigma _{ij}^{R}\left(
m_{i}^{2}\right) +\Sigma _{ij}^{L}\left( m_{i}^{2}\right) m_{j}\right] 
\\
\label{zz}
\end{equation}
It is immediate to check that
$\delta \bar{Z}_{ij}^{L}-\delta Z_{ij}^{L\dagger } \neq 0$.
This non-vanishing difference is due to the
presence of branch cuts in the self-energies that invalidate the
pseudo-hermiticity relation
$\Sigma _{ij}\left( p\right) \neq \gamma ^{0}\Sigma _{ij}^{\dagger }\left(
p\right) \gamma ^{0}$.
If we, temporally, ignore
those branch cut contributions our results reduces to the ones depicted in
\cite{Denner} or \cite{DennerSack}. In the SM these branch cuts are
generically gauge dependent as a cursory look to the appropriate integrals
shows at once. The proper consideration of the branch cuts is absolutely 
essential.

Once the off-diagonal wfr. are obtained we focus our
attention in the diagonal sector. Here, even after using the
on-shell conditions some freedom remains. We obtain
\begin{equation}
\delta m_{i}=-\frac{1}{2}Re\left\{ m_{i}\Sigma _{ii}^{\gamma L}\left(
m_{i}^{2}\right) +m_{i}\Sigma _{ii}^{\gamma R}+\Sigma _{ii}^{L}\left(
m_{i}^{2}\right) +\Sigma _{ii}^{R}\left( m_{i}^{2}\right) \right\} \,.
\label{deltam}
\end{equation}
This condition defines a mass and a width that agrees at the one-loop level
with the ones given in 
\cite{Sirlin:1991fd}. Finally,
\begin{eqnarray}
\delta \bar{Z}_{ii}^{L} &=&\Sigma _{ii}^{\gamma L}\left( m_{i}^{2}\right) -X-%
\frac{\alpha _{i}}{2}+D\,,  \notag \\
\delta \bar{Z}_{ii}^{R} &=&\Sigma _{ii}^{\gamma R}\left( m_{i}^{2}\right) +X-%
\frac{\alpha _{i}}{2}+D\,,  \notag \\
\delta Z_{ii}^{L} &=&\Sigma _{ii}^{\gamma L}\left( m_{i}^{2}\right) +X+\frac{%
\alpha _{i}}{2}+D\,,  \notag \\
\delta Z_{ii}^{R} &=&\Sigma _{ii}^{\gamma R}\left( m_{i}^{2}\right) -X+\frac{%
\alpha _{i}}{2}+D\,,  \label{zdiag}
\end{eqnarray}
\begin{eqnarray}
X &=&\frac{1}{2}\frac{\Sigma _{ii}^{R}\left( m_{i}^{2}\right) -\Sigma
_{ii}^{L}\left( m_{i}^{2}\right) }{m_{i}}\,, \\
D &=&m_{i}^{2}\left( \Sigma _{ii}^{\gamma L\prime }\left( m_{i}^{2}\right)
+\Sigma _{ii}^{\gamma R\prime }\left( m_{i}^{2}\right) \right) +m_{i}\left(
\frac{{}}{{}}\Sigma _{ii}^{L\prime }\left( m_{i}^{2}\right) +\Sigma
_{ii}^{R\prime }\left( m_{i}^{2}\right) \right)
\end{eqnarray}
Note that since $X=0$ at the one-loop level and choosing $\alpha _{i}=0$ we
obtain $\delta \bar{Z}_{ii}^{L}=\delta Z_{ii}^{L}$ and $\delta \bar{Z}%
_{ii}^{R}=\delta Z_{ii}^{R}.$ However we have the freedom to choose $\alpha
_{i}\neq 0$. This does not affect the mass terms or neutral current
couplings, but changes the charged coupling currents by multiplying the CKM
matrix $K$ by diagonal matrices. These redefinitions do not change the
physical observables provided the $\alpha _{i}$ are pure imaginary numbers.
This ambiguity corresponds in perturbation theory to the well know freedom
in phase redefinitions of the CKM matrix. Except for this last freedom, the
on-shell conditions determine one unique solution, the one presented here,
with $\bar{Z}^{\frac{1}{2}}\neq \gamma ^{0}Z^{\frac{1}{2}\dagger }\gamma
^{0} $.

\section{Gauge Invariance}

\renewcommand{\theequation}{\arabic{section}.\arabic{equation}} %
\setcounter{equation}{0}
\label{gaugeinv}
Let us briefly recapitulate where we stand at this point.
In section \ref{ward} the relation (\ref{deltaK}) relating
the CKM counterterm to a set of wfr. renormalization constants that we denoted
by $\hat{Z}$ was discussed. This 
set of renormalization constants has to fulfill the Ward identity 
(\ref{wardident}).
The diagonal character of the mass matrix is preserved along the 
renormalization process for any value of the momenta when these wfr.
constants are used.

On the other hand, in the previous section another set of wfr. constants has 
been introduced by requiring the diagonalization of the whole kinetic term
at specific values of the momenta (on-shell) as well as the corresponding unit 
residue condition.

It is important to realize that both set of wfr. constants {\em 
do not coincide}.
However, the poles are identical in both set of wfr. constants, as they
correspond, in fact, to different choices of schemes that have to render
the same set of Green functions finite. In the case of the external leg
wfr. we claim that we have no choice if we are to implement the proper
LSZ conditions. In fact, the prescription in \cite{Denner} does not achieve
the diagonalization of the absortive parts of the self-energies.
On the other hand, for the CKM counterterms one does have a choice; one 
can use for example minimal subtraction (or variations thereof), subtraction
at a given $q^2$ or whatever other method yields mass independent 
renormalization conditions. They will simply give different values of the   
renormalized entries of the CKM mixing matrix.

How can be sure that our procedure for the renormalization of the
external legs is indeed correct?
As we have seen in section \ref{ward}, gauge invariance is an issue
when dealing with CKM renormalization. Let us therefore use
gauge invariance as an additional check. We will then discover
that the presciption proposed here is the only one that provides
gauge invariant amplitudes

Let us go back to eq. (\ref{vertex}).
We know \cite{Marciano} that the combination $\frac{\delta e}{e}-\frac{%
\delta s_{W}}{s_{W}}$ is gauge parameter independent. All the other vertex
functions and renormalization constants are gauge dependent. We 
want the amplitude (\ref{vertex}) to be
exactly gauge independent ---not just its modulus--- so the gauge dependence
must cancel between all the remaining terms.

To determine the gauge dependence of the different self-energies
appearing in the external leg counterterms and the vertex 1PI Green function
we shall appeal to the so-called Nielsen identitie\cite
{Gambino,Nielsen}. We urge the interested reader to check
\cite{EMT} for details. The outcome of the analysis is that
three of the form factors
appearing in the vertex (\ref{vertex}) are by themselves gauge independent,
namely 
$ \partial _{\xi }\delta F_{L}^{\left( 2\right) }=\partial _{\xi }\delta
F_{R}^{\left( 1\right) }=\partial _{\xi }\delta F_{R}^{\left( 2\right) }=0$.
$\xi $ is the gauge-fixing parameter. We shall also see that the gauge
dependence in the remaining form factor $\delta F_{L}^{\left( 1\right) }$
cancels exactly with the one contained in $\delta Z_{W}$ and in $\delta Z$
and $\delta \bar{Z}$. Indeed the Nielsen identities lead to
\begin{equation}
\partial _{\xi }\left( M_{L}^{\left( 1\right) }\delta
F_{L}^{\left( 1\right) }\right)
=-M_{L}^{\left( 1\right)
}\partial _{\xi }\left( \delta \bar{Z}_{ir}^{uL}K_{rj}+K_{ir}\delta
Z_{rj}^{dL}+\delta Z_{W}K_{ij}\right)  \label{vertexgauge}
\end{equation}
where Eq.~(\ref{vertex}) and the gauge independence of the electric charge
and Weinberg angle has been used in the last equality. Note that Eq. (\ref
{vertexgauge}) states that the gauge dependence of the on-shell bare
one-loop vertex function cancels out the renormalization counter terms
appearing in Eq. (\ref{vertex}). This is one of the
crucial results and special care should be taken not to ignore any of the
absorptive parts ---including those in the wfr. constants. As a consequence
\begin{equation*}
\partial _{\xi }\mathcal{M}_{1}=-\frac{e}{2s_{W}}M_{L}^{\left( 1\right)
}\partial _{\xi }\delta K_{ij}\,,
\end{equation*}
so the counterterm for $K_{ij}$
must be separately gauge independent, as originally derived in \cite{Grassi}.
If one uses, for instance, the on-shell prescription wfr. constants
to determine the CKM counterterms in eq. (\ref{deltaK}), the latter will
be gauge dependent and so will be the amplitude, which is unacceptable.
On the other hand, minimal subtraction for instance is fine.

The difficulties related to a proper definition of $\delta K$ were first
pointed out in \cite{Grassi,Gambino}, where it was realized that using the
on-shell $Z$'s of \cite{DennerSack} in Eq.~(\ref{deltaK}) led to a gauge
dependent $K$ and amplitude. They suggested a modification of the on-shell
scheme based on a subtraction at $p^{2}=0$ for all flavours that ensured
gauge independence. We want to stress that the choice for $\delta K$ is not
unique and different choices may differ by gauge independent finite parts
\cite{Kniehl}.

However, this is only half the story. Assuming that a gauge independent
scheme has been used to define the $\hat{Z}$ and, accordingly, the
CKM counterterms $\delta K_{ij}$, the Nielsen identities ensure the
gauge independence of the amplitude if and only if the set of wfr. constants
derived in section \ref{legs} are used for the external legs. 
If instead of using our prescription for $\delta Z$ and $%
\delta \bar{Z}$ one makes use of the wfr. constants of \cite{Denner} to
renormalize the external fermion legs, it turns out that the gauge
cancellation dictated by the Nielsen identities does not actually take place
in the amplitude. The culprit is of course the neglect of the
absorptive parts.  Notice,
that the vertex contribution has gauge dependent absorptive parts
and they remain in the final result.

One might think of absorbing these additional terms in the counter term for $%
\delta K$. This does not quite work. Indeed one can see from 
explicit calculations that the `left-over' absorptive parts would
lead to a non-unitary CKM matrix\cite{EMT}.

It turns out that in the SM these gauge dependent absorptive parts, leading
to a gauge dependent amplitude if they are dropped, do actually cancel, at
least at the one-loop level, in the modulus of the $S$-matrix. However, in 
\cite{EMT} it is shown 
that gauge independent absorptive parts do survive
even in the modulus of the amplitude for top or anti-top decay (and only in
these cases).

\section*{Acknowledgements}

This work was triggered by many fruitful discussions with B. A. Kniehl.
The
work of D.E. and P.T. is supported in part by TMR, EC--Contract No.
ERBFMRX--CT980169 (EURODAPHNE) and contracts MCyT FPA2001-3598 and CIRIT
2001SGR-00065. It is a true pleasure to thank V.Rubakov and all the
organizing committe of the Quark 2002 Workshop for having arranged such 
a superb meeting.

\end{document}